\documentclass[pra, superscriptaddress, twocolumn, showpacs, aps, floatfix, nofootinbib]{revtex4-1}

\usepackage{amsfonts}
\usepackage{amssymb}
\usepackage{graphicx}
\usepackage{color}
\usepackage{amsmath}
\usepackage[bookmarksnumbered, bookmarks, breaklinks, linktocpage]{hyperref}

\newcommand{\Tr}        {\mathrm{Tr}}

\newcommand{\ket}[1]    {| #1 \rangle}
\newcommand{\bk}[2]     {\langle #1 | #2 \rangle}
\newcommand{\kb}[2]     {| #1 \rangle \! \langle #2 |}

\newcommand{\cS}        {{\mathcal S}}
\newcommand{\cA}        {{\mathcal A}}

\newcommand{\eend}      {\hspace{\stretch{1}}\rule{1ex}{1ex}}

{%
 \definecolor{BLACK}{gray}{0}
 \definecolor{WHITE}{gray}{1}
 \definecolor{RED}{rgb}{1,0,0}
 \definecolor{GREEN}{rgb}{0,1,0}
 \definecolor{BLUE}{rgb}{0,0,1}
 \definecolor{CYAN}{cmyk}{1,0,0,0}
 \definecolor{MAGENTA}{cmyk}{0,1,0,0}
 \definecolor{YELLOW}{cmyk}{0,0,1,0}
 }

\newcommand\dpcom[1]{}%\marginpar{\center \small $\triangleright$DP}}

\def\be{\begin{equation}}
\def\ee{\end{equation}}
\def\bea{\begin{eqnarray}}          
\def\eea{\end{eqnarray}}
\def\bi{\begin{itemize}}
\def\ei{\end{itemize}}

\usepackage{tikz}
\usetikzlibrary{arrows,shapes}

\newcommand\hocom[1]{}%\marginpar{\center \small $\triangleright$HO}}

\begin{document}

\tikzstyle{every picture}+=[remember picture]

\title{Quantum Reversibility is Relative, \\
Or \\
 Do Quantum Measurements Reset Initial Conditions?}

\author{Wojciech H. Zurek}
\affiliation{Theory Division, LANL, MS B213, Los Alamos, NM  87545}
\date{\today}

\begin{abstract} 
\noindent I compare the role of the information in the classical and quantum dynamics
by examining the relation between information flows in measurements and the ability of observers to reverse evolutions. I show that in the Newtonian dynamics reversibility is unaffected by the observer's retention of the information about the measurement outcome. By contrast---even though quantum dynamics is unitary, hence, reversible---reversing quantum evolution that led to a measurement becomes in principle impossible for an observer who keeps the record of its outcome. Thus, quantum irreversibility can result from the information gain rather than just its loss---rather than just an increase of the (von Neumann) entropy. Recording of the outcome of the measurement resets, in effect, initial conditions within the observer's (branch of) the Universe. Nevertheless, I also show that observer's friend---an agent who knows what measurement was successfully carried out and can confirm that the observer knows the outcome but resists his curiosity and does not find out the result---can, in principle, undo the measurement. This relativity of quantum reversibility sheds new light on the origin of the arrow of time and elucidates the role of information in classical and quantum physics. Quantum discord appears as a natural measure of the extent to which dissemination of information about the outcome affects the ability to reverse the measurement.
% Thus, in our quantum Universe information plays a more central role in determining what happens than in classical dynamics. 
%inconsequential, while in our quantum Universe it affects what can happen.
\end{abstract}

\maketitle

\section{Introduction}

Quantum as well as classical equations of motion are reversible. Yet, irreversibility we, observers, perceive is an undeniable ``fact of life''.  In particular, quantum measurements are famously regarded as irreversible \cite{vonN}. This irreversibility is a reason why modeling of quantum measurements using unitary dynamics is sometimes viewed as controversial. Of course, decoherence \cite{GJK+96a,Zurek03b,Schlosshauer,Zurek14} (now usually included as an essential ingredient of a fully consummated measurement process) is rightly regarded as {\it effectively} irreversible. The arrow of time it dictates can be tied to the dynamical second law \cite{Zeh,ZP}.

Our aim here is to point out that, over and above the familiar irreversibility 
%(that, {\it mutatis mutandis}, is similar in the quantum and classical settings), 
exemplified by decoherence that stems from the second law, and in contrast to the classical physics, irreversibility of an even more fundamental kind 
%that can be related to Heisenberg's indeterminacy 
arises in quantum physics in course of measurements. We shall explore it by turning ``reversibility'' from an abstract concept that characterizes equations of motion to an operationally defined property: We shall investigate when the evolution of a measured system and a measuring apparatus can be, at least in principle, reversed even if the information gained in course of the measurement is preserved (e.g., the record imprinted on the state of the apparatus pointer is copied).

This operational view of reversibility yields new insights: We shall see that reversing quantum measurements becomes impossible for an observer who retains record of the measurement outcome. This is because
%The main distinction between reversing quantum and classical evolutions is that 
the state of the measured quantum system revealed and recorded by the observer assumes---for that observer---the role reserved for the initial state in the classical, Newtonian physics. 

Consequently, clear distinction between the initial conditions and dynamics---the basis of classical physics \cite{Wigner}---is lost in a quantum setting. Indeed, quantum measurements can be reversed only when the record of the outcome is no longer preserved anywhere else in the Universe.
%This relativity of reversibility (i.e., the dependence of the ability to reverse evolution on the information at observer's disposal) sheds new light on the relativity of probabilities and, hence, on the nature of entropy as well as on the relation between objective and subjective (ontic and epistemic) aspects of quantum states.
By contrast, classical measurement can be reversed even if the record of the outcome is retained.

 Irreversibility caused by the acquisition of information in a quantum measurement has a different origin and a different character from irreversibility that follows from the second law \cite{Zeh}. There, the arrow of time -- the impossibility of reversal -- is tied to the increase of entropy, and, hence, to the loss of information. In quantum measurements irreversibility can be a consequence of the acquisition (rather than loss) of information. 

%It is related to Heisenberg-like indeterminacy: Observer who knows the outcome cannot know phases between the potential outcomes in the initial pre-measurement superposition.

This loss of the ability to reverse is {\it relative}---it depends on the information in possession of the agent attempting reversal. Thus, a friend of the observer, an agent who refrains from finding out the outcome (but can control the dynamics that led to that measurement) can, at least in principle (and in a setup reminiscent of ``Wigner's friend'' \cite{WigFr}) undo the evolution that resulted in that measurement even after he confirms that the observer had---prior to reversal---perfect record of the outcome. 

 %The separation of what happens into dynamical laws that govern the evolution and initial conditions that designate its starting point.
 Measurements re-set initial conditions relevant for observer's evolution in a manner that is tied to the choice of what is measured (as emphasized by John Wheeler \cite{Wheeler}, see Fig. 1). 
Quantum measurements (more generally, ``quantum jumps'') undermine one of the foundational principles of the classical, Newtonian dynamics:  There, consecutive measurements just narrowed down the bundle of the possible past trajectories consistent with observer's knowledge. Thus, in a classical, deterministic Universe it was always possible to imagine a single actual trajectory that fit within this bundle, and was traceable to the point marking the initial condition. This meant that evolution was reversible, an that it could be retraced---hence, reversed---using the present state of the system as a starting point into the dynamical laws and ``running the evolution backwards''.

{\begin{figure}[tb]
\begin{tabular}{l}
%\vspace{-0.15in} 
\includegraphics[width=3.5in]{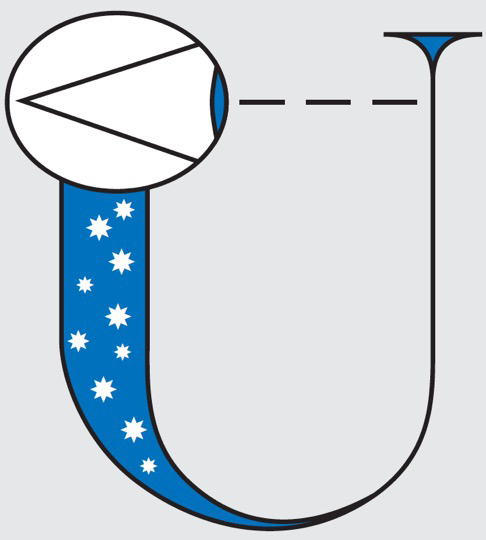}\\
\end{tabular}
\caption{An agent---an observer---within the evolving and expanding Universe carries out measurements that help define initial conditions of that Universe \cite{Wheeler}. Thus, initial conditions (at Big Bang) are determined in part by measurements carried out at present. This dramatic image (due to John Wheeler) is illustrated by the study of the ability to reverse an act of acquisition of information in this paper. 
}
\label{eye}
\end{figure}

 This idealization of a single starting point of ``my Universe''---i.e., the unique Universe consistent with the outcomes of all the past measurements at observer's disposal---is no longer tenable in the quantum setting. Quantum measurement derails evolution, resetting it onto the track consistent with its outcome. 
 
 The loss of distinction between initial conditions and dynamical laws is tied to the enhanced role of information in the quantum Universe: Information is not just a passive reflection of the deterministic trajectory dictated by the dynamics (as was imagined in the classical, Newtonian settings) but it is acquired in a measurement process that changes the state of both the measured object and of the measuring apparatus (or of an agent / observer). 
 
% The aim of this paper is to examine the role of the information in the quantum Universe through its implications for the loss of the operational reversibility in the quantum theory.

% In practice, however, this choice of what to measure is usually taken out of observer's hands by the environment that acts as a communication channel extracting the information about the preferred pointer states of the system and disseminating it, in many copies, throughout the Universe \cite{QD}. 

 We start in the next section by comparing information-theoretic prerequisites of a successful reversal in the quantum and classical case. In Section III we discuss the use of quantum discord to quantify the inability to reverse measurements. Section IV shows that another agent, a friend of the observer, can confirm that the observer is in possession of the information about the outcome in a way that does not preclude the reversal and does not reveal the outcome. This leads us to conclude that in a quantum world reversibility is indeed relative---it depends on the information in possession of the agent. Discussion and summary are offered in Section V.

 We note that much of the technical content of the paper amounts to the proverbial ``beating around the bush''. This is because the key point is ``personal'' and simple---an agent who is in possession of the information about the outcome is incapable of undoing the measurement that led to that outcome. Yet, the tools at our disposal---state vectors, density matrices, unitary  evolution operators---constrain us to discuss the measurement process ``from the outside''. And, from that external vantage point, information retained by the observer or copied into his record-keeping device plays the same role as the information acquired by the environment in course of decoherence or (especially) quantum Darwinism \cite{Zurek03b,Zurek14,QD}.
One could even say that we are stuck in the shoes of Wigner's friend \cite{WigFr}, looking at the observer ``from the outside''. 

 The ultimate message of this paper is that the observer / agent is incapable of undoing the acts of the acquisition of information, and that this inability to reverse reveals an origin of the arrow of time that is uniquely quantum and that is not dependent on the entropy increase mandated by the second law. There is of course no contradiction between the resulting arrows of time, and (as decoherence accompanies quantum measurements \cite{GJK+96a,Zurek03b,Schlosshauer,Zurek14,Zeh}) they generally appear together and point in the same direction, but they are nevertheless distinct. One way to express this difference is to note that, while our discussion is phrased in the language that presumes unitarity of evolutions, this inability to reverse may be easier to express using Bohr's ``collapse'' imagery \cite{Bohr}.

\section{Records and Reversibility}

We study operational reversibility---the ability of an observer to reverse evolution---in the classical and quantum setting. Our goal is to show that, in the quantum world, information has physical consequences that go far beyond its role in the classical, Newtonian dynamics. This illustrates the difference between the nature and function of information in quantum and classical physics. 

The key {\it gedankenexperiment} involves a measured quantum (or classical) system $\cS$ ($\bf S$), and an agent / apparatus $\cA$ ($\bf A$). The records from $\cA$ ($\bf A$) can be further copied into the memory device $\cal D$ ($\bf D$). We shall now show that presence of the copy of the record of the measurement outcomes has no bearing on the (in principle) ability to reverse a classical measurement, but precludes reversal of a quantum measurement. Thus, the pre-measurement state of the classical $\bf SA$ can be restored even when $\bf D$ knows the outcome. Such reversal is not possible for a quantum $\cal SA$ as long as $\cal D$ retains a copy of the measurement result.

%While we shall discuss the evolution of the apparatus and the memory device, $\cA$ ($\bf A$) will stand for Alice, the observer, while $\cal D$ ($\bf D$) can stand for her friend David. Both the apparatus/Alice and the memory device/David can be quantum (or classical).

It is important to emphasize the distinction between the usual discussions of reversibility (that focus on the reversibility of the equations that generate the dynamics) and our aims: Here we take for granted that it is possible to implement operators that can undo dynamical evolutions (including these leading to measurements)  in the absence of any leaks of information. Thus, in a sense, we are siding with Loschmidt in his debate with Boltzmann. For instance, we assume observer can switch the sign of the Hamiltonian that resulted in the measurement. Our aim is to shift the focus of attention from the dynamics to the role of the information observer has in implementing reversals.

\subsection{Reversing classical measurement (while keeping record of its outcome)}

We start by examining measurements carried out by a classical agent / apparatus $\bf A$ on a classical system $\bf S$. The state $\tt s$ of $\bf S$ (e.g., location of $\bf S$ in phase space) is measured (with some accuracy, but we do not need to assume perfection) by a classical $\bf A$ that starts in the ``ready to measure'' state $\tt A_0$:
$$ { \tt s A_0} \ { \stackrel {\mathfrak E_{\bf SA}} {\Longrightarrow} } \ { \tt s A_s} \eqno(1a)$$
The question we address is whether the combined state of $\bf SA$ can be restored to the pre-measurement ${ \tt s A_0}$ even when the information about the outcome is retained somewhere, e.g. copied into the memory device $\bf D$. 

The dynamics $\mathfrak E_{\bf SA}$ responsible for the measurement is assumed to be reversible and, in Eq. (1{\it a}), it is classical. Therefore, classical measurement can be undone simply by implementing ${\mathfrak E_{\bf SA}^{-1}}$ that is assumed to be at the disposal of the observer. And example of ${\mathfrak E_{\bf SA}^{-1}}$ is (Loschmidt inspired) instantaneous reversal of all velocities.

Our main point is that the reversal 
$${ \tt s A_s}  \ { \stackrel {\mathfrak E_{\bf SA}^{-1}} {\Longrightarrow} } \ { \tt s A_0} \eqno(1a')$$ 
can be accomplished even after the measurement outcome is copied onto the memory device $\bf D$:
$${ \tt s A_s D_0} \ { \stackrel {\mathfrak E_{\bf AD}} {\Longrightarrow} } \  { \tt s A_s D_s} \eqno(2a)$$
so that the pre-measurement state of $\bf S$ is recorded elsewhere (here, in $\bf D$). Above, ${\mathfrak E_{\bf AD}}$ plays the same role as ${\mathfrak E_{\bf SA}}$ in Eq. (1{\it a}). 
%(FIGURE...?)
That is, the examination of $\bf S$ and $\bf A$ separately, or of the combined $\bf SA$ will not reveal any evidence of irreversibility. After the reversal;
$${ \tt s A_s D_s} \ { \stackrel {\mathfrak E_{\bf SA}^{-1}} {\Longrightarrow} } \ { \tt s A_0 D_s} \eqno(3a)$$
 the state of $\bf SA$ is identical to the pre-measurement state, even though recording device retains the copy of the outcome.
Classical controlled-not gates provide a simple example of the claims above, as one can readily verify. 

Starting with a partly known state of the system does not change this conclusion. Thus, initial information transfer from $\bf S$ to $\bf A$:
$$ { \tt (w_s s + w_r r) A_0} \ { \stackrel {\mathfrak E_{\bf SA}} {\Longrightarrow} } \ { \tt w_s s ~A_s+ w_r r ~A_r} \eqno(4a)$$
when the system is beforehand in a classical mixture of two states $\tt {r,~s}$ with the respective probabilities $\tt {w_r,~w_s}$ can be undone---$\bf S$ and $\bf A$ will return to the initial state---even if an intermediate information transfer from $\bf A$ to $\bf D$ has occurred:
$$ { \tt ({ \tt w_s s ~A_s+ w_r r ~A_r}) D_0} \ { \stackrel {\mathfrak E_{\bf AD}} {\Longrightarrow} } \ { \tt w_s s ~A_s D_s+ w_r r ~A_r D_r} \ . \eqno(5a)$$
This is easily seen:
$$ { \tt w_s s ~A_s D_s+ w_r r ~A_r D_r}  \ {\stackrel {\mathfrak E_{\bf SA}^{-1}} {\Longrightarrow} } \  ({\tt w_s s ~D_s+ w_r r ~D_r}) {\tt {A_0}} \ . \eqno(6a)$$
In the end $\bf S$ is still correlated with $\bf D$---that is $\bf D$ has the record of the outcome of the measurement of $\bf S$ by $\bf A$. However, anyone who measures the combined state of $\bf S$ and $\bf A$ will confirm that the evolution that resulted in the measurement of $\bf S$ by $\bf A$ has been reversed. That is, the apparatus / agent $\cA$ is back in the pre-measurement state, and the system $\cS$ has the pre-measurement probability distribution over the classical microstates {\tt r,s} (even if they are still correlated with the states of the memory device {\bf D}). Thus, in classical dynamics retention of records---presence of information about the outcome of the measurement---does not preclude the ability to reverse evolutions.

\subsection{Reversing quantum measurement (can't keep the record of the outcome)}

Consider now a measurement of a quantum system $\cS$ by a quantum $\cA$:
$$ \bigl( \sum_s \alpha_s \ket s \bigr) \ket {A_0} { \stackrel {{\mathfrak U}_{\cS\cA}} {\Longrightarrow} } \  \sum_s \alpha_s \ket s \ket {A_s} \eqno(1b)$$
The evolution operator ${\mathfrak U}_{\cS\cA}$ is unitary (for example, ${\mathfrak U}_{\cS\cA}=\sum_{s,k} \kb s s  \kb {A_{k+s}}{A_k}$ with orthogonal $\{ \ket s \}$, $\{ \ket {A_k} \}$ would do the job). Therefore, evolution that leads to a measurement is in principle reversible. Reversal implemented by ${\mathfrak U}_{\cS\cA}^\dagger$ is possible, and will restore the pre-measurement state of $\cS\cA$:
$$\sum_s \alpha_s \ket s \ket {A_s} { \stackrel {{\mathfrak U}^\dagger_{\cS\cA}} {\Longrightarrow} } \ \bigl( \sum_s \alpha_s \ket s \bigr) \ket {A_0} \eqno(1b')$$
Let us however assume that the outcome of the measurement is copied before reversal is attempted:
$$ \bigl( \sum_s \alpha_s \ket s \ket {A_s} \bigr) \ket {D_0} { \stackrel {{\mathfrak U}_{\cA{\cal D}}} {\Longrightarrow} } \sum_s \alpha_s \ket s \ket {A_s} \ket {D_s} \ . \eqno(2b)$$
Here ${\mathfrak U}_{\cA{\cal D}}$ plays the same role and can have the same structure as ${\mathfrak U}_{\cS\cA}$. 

Note that Eqs. (2{\it a,b}) implement {\it repeatable} measurement / copying on the states $\{\ket s \}$, $\{ \ket {A_s} \}$ of the system and of the apparatus, respectively. That is, these states of $\cS$ and $\cA$ remain untouched by the measurement and copying processes. Repeatability implies that the outcome states $\{\ket s \}$ as well as the record states $\{ \ket {A_s} \}$ are orthogonal \cite{Zurek07,Zurek13}. This will matter in our discussion of measurements involving mixtures.

When the information about the outcome is copied, the combined pre-measurement state $\bigl( \sum_s \alpha_s \ket s \bigr) \ket {A_0}$ of $\cS\cA$ pair cannot be restored by ${\mathfrak U}_{\cS\cA}^\dagger$. 
That is:
$$ {\mathfrak U}_{\cS\cA}^\dagger \bigl( \sum_s \alpha_s \ket s \ket {A_s} \ket {D_s} \bigr) \ = \  \ket {A_0} \bigl(\sum_s \alpha_s \ket s  \ket {D_s} \bigr) \eqno(3b)$$
The apparatus is restored to the pre-measurement $\ket {A_0}$, but the system remains entangled with the memory device. On its own, its state is represented by the mixture:
$$\varrho^{\cS} = \sum _s w_{ss} \kb ss  \eqno(7) $$
where $w_{ss}=|\alpha_s|^2$. Reversing quantum measurement of a state that corresponds to a superposition of the potential outcomes is possible only providing the memory of the outcome is no longer preserved anywhere else in the Universe. 
%We shall now see that similar conclusion applies to mixed states.

\subsection{Quasiclassical case}

The special (measure zero) case when the quantum system is, prior to the measurement, in the eigenstate of the measured observable, constitutes an interesting exception to the above ``impossibility to reverse''. Then the measurement outcome:
$$ \ket s \ket {A_0} \ket {D_0} { \stackrel {{\mathfrak U}_{\cS\cA}} {\Longrightarrow} } \  \ket s \ket {A_s} \ket {D_0} \eqno(1c)$$
can be copied
$$ \ket s \ket {A_s} \ket {D_0} {\stackrel {{\mathfrak U}_{\cA\cal D}} {\Longrightarrow} } \ket s \ket {A_s} \ket {D_s} \eqno(2c)$$
and yet the evolution of $\cS\cA$ can be reversed.
$$  \ket s \ket {A_s} \ket {D_s} \  { \stackrel {{\mathfrak U}^\dagger_{\cS\cA}} {\Longrightarrow} } \  \ket s \ket {A_0}   \ket {D_s} \eqno(3c)$$
The above three equations describe evolution of quantum systems, yet they have the same structure and allow for the reversal in spite of the record retained by $\cal D$ in the same way as for the classical case (motivating the use of ``quasiclassical'' in the title of this subsection).

It is straightforward to show that the same conclusion holds for mixed states that are diagonal in the basis in which the system is measured. That is, the pre-measurement $\rho^{\cS} = \sum _s w_{ss} \kb ss $ is then identical as the post-measurement $\varrho^{\cS}$ 
%= \sum _s w_{ss} \kb ss $, 
where the "pre" and "post" are indicated in using different version of Greek ``rho''. 

This mixed quasiclassical case parallels classical Eqs. (4{\it a}-6{\it a}).  

\subsection{Superpositions of Outcomes and Measurement Reversal}

We have now demonstrated the difference between the in principle ability to reverse quantum and classical measurements. Information flows do not matter for classical, Newtonian dynamics. However, when information about a quantum measurement outcome is communicated---copied and retained by any other system---the evolution that led to that measurement cannot be reversed. Thus, from the point of view of the measurer, information retention about an outcome of a quantum measurement implies irreversibility. 

We have also examined the quasiclassical case and concluded that the presence of arbitrary superpositions in quantum theory is responsible for the irreversibility of measurements: When the considerations are restricted to such a quasiclassical set of orthogonal states, reversibility of measurements is restored. Physical significance of the phases between the potential outcomes makes quantum states vulnerable to the information leakage and prevents reversal of the evolution that led to the measurement.

This significance of arbitrary superposition was illustrated by the example of a mixture diagonal in the set of states that is left unperturbed by measurements. Measurement on a mixture that is diagonal in the same basis with which measurements correlate the state of the apparatus remains in principle reversible. Thus, in a quantum Universe where measurements are carried out only on pre-decohered systems (e.g., macroscopic systems in our Universe) and observers acquire information only about the decoherence-resistant states, one may come to believe that reversible dynamics is all there is. Of course, decoherence is an irreversible procsess, so in a sense, in our Universe, the price for this illusion of Newtonian reversibility is a massive irreversibility which is paid ``up front'', extracted by decoherence.

Presence of superpositions in correlated states of quantum systems can be quantified by quantum discord \cite{Zurek00,HV01,OZ01}. We shall now examine the relation between quantum discord and the ability to reverse measurements.

\section{Measurements of quantum mixtures, reversibility, and discord} 

The above conclusion about the impossibility to reverse quantum measurements (except for the quasiclassical case) continues to apply when the pre-measurement state of the system is a mixture diagonal in a basis that is different from the measurement basis $\{ \ket s \}$ defined by ${\mathfrak U}_{\cS\cA}=\sum_{s,k} \kb s s  \kb {A_{k+s}}{A_k}$. Thus, when the pre-measurement density matrix of the system is given by:
$$ \rho^\cS=\sum_{r,s} w_{rs} \kb r s \ , \eqno(8) $$
measurement by $\cA$ results in a combined state:
$$\bigl( \sum_{r,s} w_{rs} \kb r s \bigr) \kb {A_0} {A_0} { \stackrel {{\mathfrak U}_{\cS\cA}} {\Longrightarrow} } \sum_{r,s} w_{rs} \kb {r {A_r}} {s {A_s}} \ .\eqno(9)$$
Copying:
$$ \bigl(\sum_{r,s} w_{rs} \kb {r {A_r}} {s {A_s}}\bigr) \kb {D_0} {D_0} { \stackrel {{\mathfrak U}_{\cA{\cal D}}} {\Longrightarrow} }  \sum_{r,s} w_{rs} \kb {r {A_r}{D_r}} {s {A_s}{D_s}} \eqno(10)$$
leads to a state that exhibits quantum correlations between all three systems. Reversal of the evolution:
$$ \sum_{r,s} w_{rs} \kb {r {A_r}{D_r}} {s {A_s}{D_s}} { \stackrel {{\mathfrak U}_{\cS\cA}^\dagger} {\Longrightarrow} } \kb {A_0} {A_0} \sum_{r,s} w_{rs} \kb {r {D_r}} {s {D_s}} \eqno(11)$$
 that acts purely on the $\cS \cA$ pair restores only the pre-measurement state of the apparatus, but not the state of the system,  
 $$\varrho^{\cS} = \sum _s w_{ss} \kb ss =\Tr_{\cal D}  (\sum_{r,s} w_{rs} \kb {r {D_r}} {s {D_s}}) \ ,
 \eqno(12)$$ 
 as the reduced density matrix of the system
 is now---unlike the pre-measurement $\rho^\cS$, Eq. (8)---diagonal in the measurement basis $\ket s$. 
 
Thus, in contrast to the classical case, acquiring and communicating information about quantum systems matters: Reversibility of the global dynamics is not enough. Presence of a copy of the information (that did not matter in the classical case) precludes the possibility of implementing local reversals.

The information-theoretic price---the extent of irreversibility---can be quantified by $ \Delta H$, the difference in entropy between the pre-measurement and post-measurement density matrices;
$$ \Delta H = -\sum_s w_{ss}\lg w_{ss} + \Tr \rho_\cS \lg  \rho_\cS = H( \varrho_\cS ) - H(\rho_{\cS}) \ . \eqno(13)$$
We shall now show that this entropy increase caused by copying coincides with the quantum discord \cite{Zurek00,HV01,OZ01} in the correlated post-measurement state $\sum_{r,s} w_{rs} \kb {r {A_r}} {s {A_s}}$ of the system and the apparatus. This suggests that vanishing of discord may be a condition for the reversibility undisturbed by copying. 
%We shall now explore this discord-reversibility connection in greater detail.

\subsection{Introducing quantum discord}

Discord is the difference between the mutual information defined by the symmetric equation that involves von Neumann entropies of the two systems separately and jointly:
$$ I(\cS:\cA) = H_\cS + H _\cA - H_{\cS\cA} \ , \eqno(14)$$
where $H_X=-\Tr \rho_X \lg \rho_X$, and the asymmetric definition of mutual information $J(\cS;\cA)_{\cA|{\{\ket {A_k} \}} }$.
%where $H_\cS,~ H _\cA, ~H_{\cS\cA}$ are the individual and joint von Neumann entropies.

The asymmetric version of mutual information obtains from the joint entropy when it is expressed in terms of the conditional entropy: 
$$H_{{\cS\cA}|{\{\ket {A_k} \}  }} = H_{\cS|\cA{\{\ket {A_k} \}  }}+ H_{\cA|{\{\ket {A_k} \} }} \ ,  \eqno(15)$$
where we have assumed that the measurements were performed on $\cA$ in the basis $\{\ket {A_k} \}$. Thus, $H_{\cA|{\{\ket {A_k} \} }}$ is the entropy computed using probabilities of states $\{\ket {A_k} \}$, and $H_{\cS|\cA{\{\ket {A_k} \}  }}$ is the conditional entropy one still has after the outcomes of measurement on $\cA$ in the basis $\{\ket {A_k} \}$ are known.
%The individual terms is the above equation are averages over the properly normalized density matrices corresponding to the outcomes of measurement on $\cA$ 

In the classical setting, when Shannon entropies are computed from classical probabilities, analogous two expressions for the joint entropy coincide
%, $H_{{\bf S A}|{\tt A}_k} =H_{\bf  SA}$ 
\cite{CoverThomas}.
However, in the quantum setting, possible post-measurement states---hence, conditional information---have to be defined with respect to the basis set characterizing the measurement that is carried out on one of the two systems (here $\cA$) in order to gain partial information about the other (here $\cS$). 
Using this basis-dependent joint entropy $H_{{\cS\cA}|{\{\ket {A_k} \}  }}$ in Eq. (14) instead of $H_{\cS\cA} $ one gets an asymmetric expression for mutual information:
$$ J(\cS;\cA)_{\cA|{\{\ket {A_k} \} }}=H_\cS + H_\cA - (H_{\cS|{\cA \{\ket {A_k} \}  }}+ H_{\cA|{\{\ket {A_k} \} }}) \ . \eqno(16) $$

Discord is the difference between the symmetric and asymmetric formulae for mutual information\footnote{There are subtleties in the definition of the discord.  Definition given here is the so-called thermal discord or one-way deficit. It differs from the ``original'' discord defined in \cite{Zurek00,HV01,OZ01}. A brief discussion in the context of Maxwell's demon can be found in \cite{Zurek03}. More extensive discussions of discord and related measures are also available \cite{Modi,Sen}. We note that appearance of discord in the correlated $\cS\cA$ state can be traced \cite{Girolami} to the presence of quantum coherence in the states of $\cS$.}:
$$\delta(\cS:\cA)_{\cA|{\{\ket {A_k} \} }}=I(\cS:\cA)-J(\cS;\cA)_{\cA|{\{\ket {A_k} \} }} \ , \eqno(17a) $$
or;
$$\delta(\cS:\cA)_{\cA|{\{\ket {A_k} \} }}=(H_{\cS|{\cA \{\ket {A_k} \}  }} + H_{\cA|{\{\ket {A_k} \} }})-H_{\cS\cA}  \ . \eqno(17b)$$
When the two systems are classical (so that their states can be completely described by probabilities) the two definitions of the mutual information coincide, and quantum discord disappears---it is identically equal to zero. 
In the quantum domain probabilities usually do not suffice, and the two expressions for the mutual information differ. 
%The resulting discord depends on the basis $\{\ket {A_k} \}$ in which the measurement is carried out. 

In the case we have considered above the system was in a mixed state, but the initial state of the apparatus was pure, and the measurement that correlated $\cS$ with $\cA$ was unitary, so that $H_{\cS\cA}=H(\rho_\cS )$. Moreover, $H_{\cS|{\cA\{\ket {A_s} \}  }}=0$ (as a measurement of $\cA$ with the result $\ket {A_k} $ reveals the corresponding pure states of $\cS$) and $H_{\cA|{\{\ket {A_s} \}}}=H(\varrho_\cS )$ (as the entropy of $\cA$ is, after it correlates with $\cS$ computed from the probabilities $w_{ss}$ and equals $-\sum_s w_{ss}\lg w_{ss}$). Consequently, the entropy increase $\Delta H$ of Eq. (13) is indeed equal to the discord in the post-measurement (but pre-copying) state of $\cS\cA$.

%, as the initial state of the apparatus was pure and the evolution $$ was unitary.

%When the joint density matrix of $\cS\cA$ is diagonal in the measurement basis $\{\ket {A_k} \}$, discord will vanish. However, this will not be generally the case -- such a basis may not exist.
%In our case the copying will be carried out on $\cA$, and it need not be exhaustive---the apparatus can be macroscopic, so the information about $\cS$ can be stored in the subspaces if $\cA$. 
%The apparatus is usually macroscopic (but still quantum) system. Therefore, the will correspond to subspaces of its Hilbert space -- subspaces that provide support of the density matrices. As we have noted earlier, these subspaces should be orthogonal to enable repeated copying \cite{Zurek13}. 

\subsection{Reversibility and quantum discord}

We now consider a general case, where the pre-measurement density matrices $\rho^\cS$, $\rho^\cA$ and the post-measurement $\rho^{\cS\cA}$ can all be mixed. The evolution that leads to the measurement is still unitary ${{\mathfrak U}_{\cS\cA}}$. And we still assume that the apparatus should obtain and retain at least an imperfect record of the system. 
That is, there should be states $\{\rho_s^\cS\}$ of the systems that leave imprints on the state of the apparatus:
$$ \rho_s^\cS \rho_0^\cA  \  { \stackrel {{\mathfrak U}_{\cS\cA}} {\Longrightarrow} } \ \rho_s^{\cS\cA} \ . \eqno(18)$$
An initial mixture of $\{\rho_s^\cS\}$ will evolve, by linearity, into the corresponding mixture of the outcomes. 
$$ \sum_s p_s \rho_s^\cS \rho_0^\cA  \  { \stackrel {{\mathfrak U}_{\cS\cA}} {\Longrightarrow} } \  \sum_s p_s \rho_s^{\cS\cA} = \rho^{\cS\cA} \ . \eqno(19)$$
The correlation could be imperfect (i.e., one might only be able only infer some information about some of the $\{\rho_s^\cS\}$ from $\cA$). 
\hocom{A simple way to express this is to demand that the quantum mutual information:
$$ I(\cS:\cA) = H_\cS + H _\cA - H_{\cS\cA} $$
should be non-zero where $H_\cS,~ H _\cA, ~H_{\cS\cA}$ are the individual and joint von Neumann entropies computed using the post - measurement reduced density matrices $\rho^\cS, \rho^\cA$, and $\rho^{\cS\cA}$ of $\cS\cA$ from which they obtain. }

Copying involves interaction of $\cA$ and $\cal D$. As before, we enquire under what circumstances transfer of information about $\cS$ via $\cA$ to $\cal D$ does not preclude reversal, so that the evolution generated by ${\mathfrak U}_{\cS\cA}^\dagger$ restores the pre-measurement state of $\cS\cA$ in spite of the correlation with $\cal D$ established by:
$$ \sum_s p_s \rho_s^{\cS\cA} \kb {D_0} {D_0} \ { \stackrel {{\mathfrak U}_{\cA\cal D}} {\Longrightarrow} } \  \sum_s p_s \rho_s^{\cS\cA} \kb {D_s} {D_s} = \rho^{\cS\cA\cal D} \eqno(20)$$
To allow for reversal the state of ${\cS\cA}$ must not be affected by the copying. That is, 
$$\varrho^{\cS\cA}=\Tr_{\cal D} \rho^{\cS\cA\cal D}=\rho^{\cS\cA}\ , \eqno(21)$$ 
where $\rho^{\cS\cA}$ and $\varrho^{\cS\cA}$ are the density matrices before and after the copying operation.
This is a density matrix version version of the ``repeatability condition'' (see \cite{Zurek07,Zurek13}): Copying can be repeated (since the ``original'' remains unchanged), and we shall see that this repeatability leads to similar consequences---to the orthogonality of the records that can be copied. 

%Furthermore, repeatability implies that we could take the copies in $\cal D$ to correspond to orthogonal states (i.e., $\bk {D_r} {D_s} = \delta_{rs}$, see \cite{Zurek07} for details). We shall actually not assume such orthogonality below (as it is easier to follow some of the reasoning without relying on this assumption).

Unitarity of ${\mathfrak U}_{\cA\cal D}$ is 
%(as in the other cases of copying) 
responsible for our next result. Unitary evolutions preserve Hilbert-Schmidt norm. Therefore, 
$$\sum_{r,s} p_r p_s \Tr \rho_r^{\cS\cA} \rho_s^{\cS\cA} = \sum_{r,s} p_r p_s \Tr \rho_r^{\cS\cA} \rho_s^{\cS\cA} |\bk {D_r} {D_s}|^2 \ . \eqno(22)$$
The overlap of the copy states in $\cal D$ is non-negative and bounded, $0 < |\bk {D_r} {D_s}|^2 \le 1$. Therefore, there are only two ways to satisfy this equality: Either $|\bk {D_r} {D_s}|^2=1$ (i.e., there is no copy!), or 
$$p_r p_s \Tr \rho_r^{\cS\cA} \rho_s^{\cS\cA}=0 \ . \eqno(23)$$ 
For the non-trivial case when $p_r p_s>0$ and $r \neq s$ this leads to; 
$$\Tr \rho_r^{\cS\cA} \rho_s^{\cS\cA}=0 \eqno(24)$$
as a necessary condition to allow for copying that does not interfere with the possibility of the reversal.  

Indeed, when (as we have assumed) copying evolution operator ${\mathfrak U}_{\cA\cal D}$ involves only $\cA$ and $\cal D$, we can repeat the above reasoning starting with the reduced density matrix of $\cA$ alone and demanding that it is untouched by the copying operation:
$$\varrho^{\cA}=\Tr_{\cS \cal D} \rho^{\cS\cA\cal D}=\rho^{\cA}\ . \eqno(25)$$
(Clearly, if copying were to affect density matrix of $\cA$, it would affect also $\rho^{\cS\cA}$, so Eq. (21) cannot not be satisfied unless Eq. (25) holds.)

In the end we will conclude that repeatability is not ruled out by retention of the copies of the outcomes providing that:
$$p_r p_s \Tr \rho_r^{\cA} \rho_s^{\cA}=0 \ . \eqno(26)$$ 
For the non-trivial case when $p_r p_s>0$ this implies orthogonality of the records:
$$\Tr \rho_r^{\cA} \rho_s^{\cA}=0 \eqno(27)$$ 
as a necessary condition to allow for copying of the information from $\cA$ that does not interfere with the possibility of the reversal. 

To assure that copying will indeed leave $\varrho^{\cS\cA}$ unchanged, we need satisfy the same condition that selects pointer states \cite{Zurek81,Zurek82}: The unitary ${\mathfrak U}_{\cA \cal D}$ that produces copies must commute with the pre-copying $\varrho^{\cS\cA}$ to leave it unaffected. 
%As  ${\mathfrak U}_{\cA \cal D}$ couples only with the apparatus, this means that the record states of $\cA$. 
This will be the case when the Hamiltonian $\bf H_{\cA \cal D}$  that generates ${\mathfrak U}_{\cA \cal D}$ commutes with the pointer observable of $\cA$---with the apparatus observable that keeps the records of the state of the system. This pointer observable will have in general degenerate eigenstates---eigenspaces that serve (within the apparatus Hilbert space) as a ``one leg'' of the support of the density matrices  $\rho_s^{\cS\cA}$. Orthogonality of the record states of $\cA$ implies zero ``one way'' discord in the basis corresponding to these pointer eigenspaces. 

We note that there is an important difference between Eqs. (23,~24) and Eqs. (26,~27) we have derived. They rely on different assumptions: Eqs (26,~27) are ``local'' -- they focus on the content of the records in the apparatus alone, and demand distinguishabiility (orthogonality) of its states. This focus is justified by the nature of the copying interaction---it involves only $\cA$ and $\cal D$, so only the records in $\cA$ are relevant. By contrast, Eqs. (23,~24) could be satisfied equally well by orthogonality of local state of $\cS$ alone or, indeed, of the global states of $\cS\cA$. In other words, when one can access the composite system $\cA\cS$, the condition that allows for reversible copying can be satisfied by the global state even when it is not met by the record states of $\cA$ alone \cite{Zurek13}. Our next goal is to consider effects of such more global copying operations.

\section{Knowing of the record but not the outcome}

Immediately above, in Eqs. (25-27), we have insisted that the orthogonality condition $\Tr \rho_r^{\cS\cA} \rho_s^{\cS\cA}=0$ should be satisfied ``in the apparatus'', that is, that the apparatus eigenspaces that correspond to the records should be orthogonal. This insistence stemmed from the fact that the copying evolution ${\mathfrak U}_{\cA\cal D}$ coupled only to $\cA$. However, one can imagine a situation where ${\mathfrak U}_{(\cS\cA)\cal D}$ couples $\cal D$ to a global observable of $\cS\cA$. In that case, one might be able to find out that $\cA$ ``knows'' the outcome -- the state of $\cS$ -- without actually finding out the outcome. 
%This would be the case when, for instance, the eigenstates of the global observable of $\cS\cA$ are entangled. 

The simplest such example is afforded by a one qubit apparatus that measures a one qubit system. 
%One could then make copies in e.g. the Bell basis of $\cS\cA$. 
The correlated---entangled---state of the two is then simply:
$$ \ket {\psi_{\cS\cA}} = a_\uparrow \ket {\uparrow A_\uparrow} + a_\downarrow \ket {\downarrow A_\downarrow} \eqno(28)$$
in obvious notation. Agent $\cal D$ can then detect presence of the correlations established when $\cS$ and $\cA$ interacted.

We now consider two operators that can confirm the existence of the correlation between $\cS$ and $\cA$. The first such operator, when measured, would establish whether the states of $\cS$ and $\cA$ are correlated in the basis (here $\{ \ket \uparrow, \ket \downarrow\}$) in which the measurement was carried out:
$${\bf \hat A} = y_\uparrow \kb {\uparrow A_\uparrow} {\uparrow A_\uparrow} + y_\downarrow \kb {\downarrow A_\downarrow} {\downarrow A_\downarrow} $$
$$ + n \kb {\uparrow A_\downarrow} {\uparrow A_\downarrow} + n' \kb {\downarrow A_\uparrow} {\downarrow A_\uparrow} \ . \eqno(29) $$
The detection of either of the $y$ eigenvalues would imply a successful measurement (while either of $n$ eigenvalues would signify error). Moreover, when $y_\uparrow= y_\downarrow=y$, such measurement would reveal consensus without betraying the actual outcome. Thus, agent $\cal D$---friend of the observer---could confirm the success of the measurement, but the evolution that led to the measurement can be be still undone. 

This is ``relative reversibility''---the evolution that led to measurement can be at least in principle undone by an agent who can confirm that the measurement was successful providing he does this without finding out the outcome. When $y_\uparrow \neq y_\downarrow$, the measurement by $\cal D$ would correlate his state with the outcome, and the reversal would become impossible. 

%We note that $\bf \hat A$ will ``work'' even after the apparatus was decohered by the environment in the pointer basis $\{\ket {A_\uparrow}, \ket {A_\downarrow}$. 

An alternative confirmation of a successful $\cS\cA$ measurement can be accomplished by detecting entanglement in $\ket {\psi_{\cS\cA}}$. Bell operator:
$$
{\bf \hat B} = b^+_= \kb {\beta^+_=}{\beta^+_=} + b^-_= \kb{\beta^-_=}{\beta^-_=} + 
b^+_{\neq} \kb {\beta^+_{\neq}}{\beta^+_{\neq}} + b^-_{\neq} \kb {\beta^-_{\neq}}{\beta^-_{\neq}} \eqno(30)$$
can be used for this purpose. 
Above, subscripts ``='' and ``$\neq$'' stand for ``parallel'' and ``antiparallel'', and the Bell eigenstates are;
$$\ket {\beta^\pm_=} = \ket {\uparrow A_\uparrow} \pm \ket {\downarrow A_\downarrow} \ ; \eqno(31a)$$
$$\ket {\beta^\pm_{\neq}} = \ket {\uparrow A_\downarrow} \pm \ket {\downarrow A_\uparrow} \ . \eqno(31b) $$
Detection of either $b^+_=$ or $b^-_=$ implies successful measurement. However, unless $b^+_= = b^-_=$, measurement will also reveal phases between the outcome states, and (unless $\ket {\psi_{\cS\cA}}$ happens to be one of the above Bell states) it will result in decoherence in the Bell basis (and, hence, prevent reversal).

It is interesting to note that when one imposes degeneracy that enables reversal on either $\bf \hat A$ or $\bf \hat B$, eigenstates of these two operators coincide. The resulting consensus operator is given by:
$${\bf \hat C} = y (\kb {\uparrow A_\uparrow} {\uparrow A_\uparrow} + \kb {\downarrow A_\downarrow} {\downarrow A_\downarrow})$$
$$ + n  (\kb {\uparrow A_\downarrow} {\uparrow A_\downarrow} +  \kb {\downarrow A_\uparrow} {\downarrow A_\uparrow}) \eqno(32)$$
Thus, by measuring $\bf \hat C$ one can confirm that $\cA$ ``knows the outcome'' without impairing the possibility of reversal.

\section{Discussion}

Our results shed new light both on the relation between quantum and classical and on the role of information in measurements. So far we have mainly emphasized their relevance for the distinction between quantum and classical physics. To re-state briefly the main conclusion, retention of information about classical states has no bearing on the in principle ability to reverse classical evolution that leads to measurement, but it precludes reversing quantum measurements (with the exception of the quasiclassical case). Thus, information plays a far more important role in quantum Universe than it used to play in classical physics.

This operational view of reversibility yields new insights:
%We shall conclude that while dynamics leading to classical measurements can be reversed even if the information gained in course of the measurement is retained (e.g., because a copy of the outcome is made), quantum measurements 

(i) In quantum physics irreversibility in course of  measurements need not be blamed solely on decoherence, but is caused by observer's acquisition of the data about the system. Observer who retains record of the outcome cannot restore the pre-measurement states of both the system $\cS$ and the apparatus $\cA$. So, from observer's point of view, while classical measurements can be undone, quantum measurements are fundamentally irreversible.

(ii) Acquisition of information results in decrease of the von Neumann entropy of the system. Therefore, this aspect of irreversibility of measurements is not a consequence of the second law. Yet, while observer can take advantage of this (apparent!) violation of the second law, he cannot reverse measurement on his own. 

(iii) However, observer's friend (who knows about the measurement, but not its outcome) can, in principle, induce such a reversal providing there is no copy of the record of the outcome left anywhere.

%she does not find out the outcome (e.g., does not make a copy of its record).

%(iv) Inability to reverse is ultimately tied to Heisenberg's indeterminacy: Observer cannot simultaneously know the outcome and reverse the evolution of the global state (that represents superposition of all the outcomes consistent with the pre-measurement states and the phases between them). 

Our discussion calls for a re-consideration of the nature and origin of the initial conditions in quantum physics. Distinction between the laws that dictate evolution of the state of a system and initial conditions the define its starting point dates back to Newton \cite{Wigner}. This clean separation is challenged by quantum measurements. Seen from the inside, by the observer, measurement re-sets initial conditions. Acquisition of information simultaneously redefines the state of the observer and observer's branch of the universal state vector. From then on, observer will exist within the Universe he helped define (see Fig. 1). On the other hand, observer's friend will---for as long as he does not find out what the observer found out---live in a Universe where the initial condition is the pre-measurement state with a coherent superposition of all the potential outcomes.

Familiar ``paradox'' of Wigner's friend offers an interesting setting for this discussion. Wigner speculated  \cite{WigFr} (following to some extent von Neumann \cite{vonN}) that ``collapse of the wavepacket'' may be ultimately precipitated by consciousness. The obvious question is, of course, ``how conscious should the observer be''. 

The answer suggested by our discussion is that---if the evidence of collapse is the irreversibility of the evolution that caused it---retention of the information suffices. Thus, there is no need for ``consciousness'' (whatever that means): Record of the outcome is enough. On the other hand, observer conscious of the outcome certainly retains its record, so being conscious of the result suffices to preclude the reversal---to make the ``collapse'' irreversible. Quantum Darwinism \cite{QD} traces emergence of the objective classical reality to the proliferation of information throughout the environment. Our discussion of the consequences of retention of information for reversibility is clearly relevant in this context, although its detailed study is beyond the scope of this paper.

%There is not contradiction between these two statements. The {\it existential interpretation} recognized this dual role of quantum evolution in the context of decoherence \cite{Zurek03b}. 

This research was supported by the Department of Energy via LDRD program in Los Alamos, and, in part, by the Foundational Questions Institute grant ``Physics of What Happens''.

\end{document}